\documentclass[aip,jcp,reprint,amsmath,amssymb]{revtex4-1}
\usepackage{graphicx}
\usepackage{dcolumn}
\usepackage{bm}
\usepackage[utf8]{inputenc}
\usepackage[T1]{fontenc}
\usepackage{mathptmx}
\usepackage{etoolbox,float}
\usepackage{booktabs}
\usepackage[version=3]{mhchem}

\makeatletter
\def\@email#1#2{%
 \endgroup
 \patchcmd{\titleblock@produce}
  {\frontmatter@RRAPformat}
  {\frontmatter@RRAPformat{\produce@RRAP{*#1\href{mailto:#2}{#2}}}\frontmatter@RRAPformat}
  {}{}
}%
\makeatother

\begin{document}
\preprint{}

\title{PBr$_3$ Adsorption and Dissociation on the Si(100) Surface}

\author{Vladimir M. Shevlyuga}
\affiliation{Prokhorov General Physics Institute of the Russian Academy of Sciences, Vavilov str. 38, 119991 Moscow, Russia}

\author{Yulia A. Vorontsova}
 \email{pavlova@kapella.gpi.ru}
\affiliation{Prokhorov General Physics Institute of the Russian Academy of Sciences, Vavilov str. 38, 119991 Moscow, Russia}
\affiliation{HSE University, Myasnitskaya str. 20, 101000 Moscow, Russia}

\author{Tatiana V. Pavlova}
 \email{pavlova@kapella.gpi.ru}
\affiliation{Prokhorov General Physics Institute of the Russian Academy of Sciences, Vavilov str. 38, 119991 Moscow, Russia}
\affiliation{HSE University, Myasnitskaya str. 20, 101000 Moscow, Russia}

\date{\today}

\begin{abstract}

The adsorption of \ce{PBr3} on the Si(100)-2$\times$1 surface was studied by scanning tunneling microscopy (STM) and density functional theory (DFT). The \ce{PBr3} molecule completely dissociates on the Si(100) surface at room temperature into P and Br atoms. In most cases, the dissociated molecule was observed in STM on three neighboring Si dimers. DFT calculations confirm that the \ce{PBr3} molecule can completely dissociate at room temperature. After annealing the sample to 400$^{\circ}$C, phosphorus is incorporated into silicon, as evidenced by the Si atoms ejected to the surface. These findings are useful for the insertion of individual phosphorus atoms into silicon by \ce{PBr3} adsorption through a halogen mask.

\end{abstract}

\maketitle

\section{\label{sec:intro}Introduction}

Phosphine is currently the main precursor molecule for doping silicon with phosphorus for semiconductor devices, including doping of the Si(100) surface with near-atomic precision to create quantum devices \cite{2012Fuechsle, 2019He, 2022Kiczynski, 2022Wang}. Due to technological importance, the mechanism of phosphorus incorporation into silicon by phosphine adsorption has now been studied in detail. The \ce{PH3} molecule adsorbed on the Si(100) surface dissociates at room temperature \cite{2005Warschkow, 2016Warschkow}. For complete dissociation of phosphine, three adjacent silicon dimers are required \cite{2004Wilson, 2016Warschkow}. Subsequent heating of the sample with adsorbed phosphine leads to the incorporation of phosphorus atoms into surface layer, as evidenced by Si atoms ejected from the substrate \cite{2003Schofield}.

To dope the Si(100) surface with phosphorus with near-atomic precision, a hydrogen monolayer patterned in a scanning tunneling microscope (STM) is used as a mask \cite{2003Schofield}. An alternative to the hydrogen mask on the Si(100) surface can be a halogen monolayer \cite{2018Pavlova, 2019Dwyer, 2020Pavlova, 2022Pavlova}. The halogen monolayer may have a number of advantages, since a halogenated silicon surface has a different reactivity compared to a hydrogenated one. In particular, calculations show that the halogen monolayer can better protect the Si(100) surface against parasitic incorporation of impurities from halogen-containing precursor molecules \cite{2021Pavlova}. Recently, the interaction of \ce{BCl3} with chlorinated and hydrogenated Si(100) surfaces \cite{2020Silva-Quinones, 2021DwyerACS} and \ce{AlCl3} with a clean Si(100) surface \cite {2021Radue} has been studied.

Here, as a source of phosphorus for silicon doping, we consider the \ce{PBr3} molecule, which should be compatible with a halogen mask on the Si(100) surface. We studied the \ce{PBr3} adsorption on the clean Si(100) surface at room temperature in STM. Fragments of the dissociated \ce{PBr3} molecule were identified using density functional theory (DFT).
After \ce{PBr3} adsorption, phosphorus was incorporated into the silicon surface by annealing the sample to 400$^{\circ}$C.

\section{Methods}

\subsection{Experimental methods}

The experiments were carried out in an ultra-high vacuum (UHV) system with a base pressure of 5$\times$10$^{-11}$\,Torr. The STM measurements were performed with GPI CRYO (SigmaScan Ltd.) operated at 77\,K. B-doped Si(100) samples (1\,$\Omega$\,cm) were used. The Si(100) sample was prepared by outgassing the wafer at 870\,K for several days in UHV followed by flash-annealing at 1470\,K. After the last flash heating, the sample was allowed to cool down to room temperature for 30 min. The \ce{PBr3} adsorption was carried out at a partial pressure of $3 \cdot 10^{-10}$\,Torr for two minutes at room temperature. To identify Br atoms from the dissociated \ce{PBr3} molecule, we carried out the \ce{Br2} adsorption on the clean Si(100) surface at a partial pressure of $1.5\cdot 10^{-10}$\,Torr for ten seconds at room temperature. To remove phosphorus from the surface for the next experiment, the sample was heated at 1170\,K overnight \cite{2005Brown}. We used mechanically cut Pt-Rh and Pt-Ir tips and electrochemically etched polycrystalline W tips. The voltage ($U_s$) was applied to the sample. All STM images were processed using the WSXM software \cite{WSXM}.

\subsection{Computational methods}

The spin-polarized DFT calculations were performed with the Perdew-Burke-Ernzerhof (PBE) functional \cite{1996Perdew} as implemented in the Vienna \textit{ab initio} simulation package (VASP) \cite{1996Kresse, 1999Kresse}. The kinetic-energy cutoff of the plane wave basis was set to 350\,eV. The silicon surface was modeled by an eight layer slabs consisting of 6$\times$6 supercell and a 14\,{\AA} vacuum gap. Phosphorus and bromine atoms were placed on the top Si(100) surface with the 2$\times$1 structure, whereas hydrogen atoms saturate the dangling bonds at the bottom Si surface. The bottom three Si layers were frozen at their bulk positions, while the coordinates of other atoms were fully relaxed until the residual forces were smaller than 0.01\,eV/\,{\AA}. Brillouin zone integrations were done using a 3$\times$3$\times$1 k-point grid. We studied the convergence of the bromine adsorption energy on the Si(100) surface as a function of the number of substrate layers, cutoff energy, k-points, and vacuum gap. The adsorption energies were calculated as the difference between the total energy of the surface with an adsorbed molecule and the total energy of the clean surface and the molecule in vacuum. STM images were generated in the HiveSTM program \cite{2008Vanpoucke} using the Tersoff-Hamann approximation \cite{1985Tersoff}.

The activation barriers were calculated by using the nudged elastic band (NEB) method \cite{1998NEB}. In this case, the 4$\times$4 supercell was used, and the criterion of convergence for residual forces was reduced to 0.03\,eV/\,{\AA}. We used five images between two endpoints. The reaction rate was estimated using the Arrhenius equation:
\begin{equation}
\nu = \nu_0 \exp \Bigl( - \frac{E_{act}}{k_B T} \Bigr), \label{eq:1}
\end{equation}
where $E_{act}$ is the activation energy, $k_B$ is the Boltzmann constant, $T$ is the temperature. Attempt frequency $\nu_0$ was determined by the Vineyard equation \cite{1957Vineyard}:
\begin{equation}
\nu_0 =  \prod\limits_{n=1}^{3N} \nu^{IS}_n \Bigl/ \prod\limits_{n=1}^{3N-1} \nu^{TS}_n , \label{eq:2}
\end{equation}
where $\nu^{IS}_n$ and $\nu^{TS}_n$ are the real-valued vibrational frequencies in the initial (IS) and transition (TS) states, respectively. The number N of atoms for which the vibration frequencies were calculated were limited by the nearest neighbors of the migrating atoms. The transition state was verified by the presence of a single imaginary frequency. The frequencies were calculated using the density functional perturbation theory (DFPT), as implemented in VASP.

\section{Results and discussion}

\begin{figure}[t!]
\begin{center}
\includegraphics[width=\linewidth]{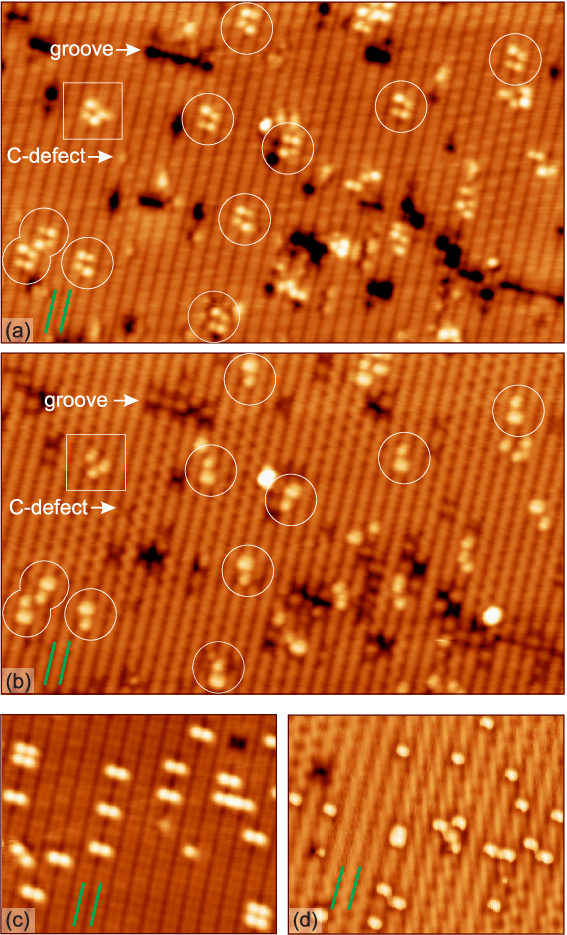}
\caption{\label{STM_large} (a) Empty ($U_s =+2.2$\,V) and (b) filled ($U_s =-3.5$\,V) state STM images (30.1$\times$18.2\,nm$^2$, I$_t$ = 1.5\,nA) of the Si(100) surface after \ce{PBr3} adsorption at room temperature. The most frequently observed object on the \ce{PBr3}-dosed surface is marked with a white circle. One of the other observed objects is indicated by a white square. (c) Empty ($U_s =+2.5$\,V, I$_t$ = 2.2\,nA) and (d) filled ($U_s =-3.5$\,V, I$_t$ = 1.3\,nA) state STM images (11.1$\times$8.8\,nm$^2$) of the Si(100) surface after \ce{Br2} adsorption. Two different surface areas are shown in (c) and (d). The gaps between rows of Si dimers are marked with green lines.}
\end{center}
\end{figure}

Figures~\ref{STM_large}a,b shows empty and filled states STM images of the Si(100) surface after \ce{PBr3} dosing at room temperature. After \ce{PBr3} adsorption, the most frequently observed object has a trapezoidal shape with four protrusions in the empty state STM image (Fig.~\ref{STM_large}a), while only three protrusions are visible in the filled states STM image (Fig.~\ref{STM_large}b). Such objects accounted for about half of all observed objects. Unfortunately, the number of the obtained STM images with atomic resolution are not sufficient for qualitative statistics.

The main defects on the surface are C-defects \cite{2015Smith} (see Supplementary Material, SM) and grooves, which were also present before \ce{PBr3} adsorption. We associate the grooves with the vacancy line defects observed in Ref. \cite{2005Brown}, where it was shown that the vacancy line defects are caused by stress due to the presence of phosphorus near the surface. As in Ref. \cite{2005Brown}, we observed a decrease in the number of the vacancy line defects when the sample was heated to 1170\,K, at which phosphorus, rather than metal impurities, desorbed. In our case, P-doping occurs due to repeated experiments on the \ce{PBr3} adsorption on the same sample, in contrast to Ref. \cite{2005Brown}, in which a heavily P-doped sample was used.

To distinguish Br atoms from P atoms, we dosed Si(100) with low exposure \ce{Br2}. Bromine is visualized in the STM images as protrusions at sufficiently high positive ($U_s>2$ ) and negative ($U_s<-3$) voltages. Each Br atom forms a bond with one Si atom and is located above this atom. At a low coverage, two Br atoms mainly occupy one Si dimer (Fig.~\ref{STM_large}c, d).

To identify features observed in STM images after \ce{PBr3} dosing, we simulated STM images of the \ce{PBr3} molecule adsorbed on the surface, as well as all possible fragments after \ce{PBr3} dissociation (see SM). The calculated structures with \ce{PBr3}, \ce{PBr2}+Br, \ce{PBr}+2Br, and P+3Br are denoted as pbr3-n, pbr2-n, pbr-n, and p-n, respectively, where n is the number of the structure (a smaller number n corresponds to a more stable structure). We considered the dissociation of the \ce{PBr3} molecule into different fragments on no more than three dimers in one row, because the most frequently observed object occupies three neighboring dimers.

\begin{figure}[h!]
\begin{center}
\includegraphics[width=\linewidth]{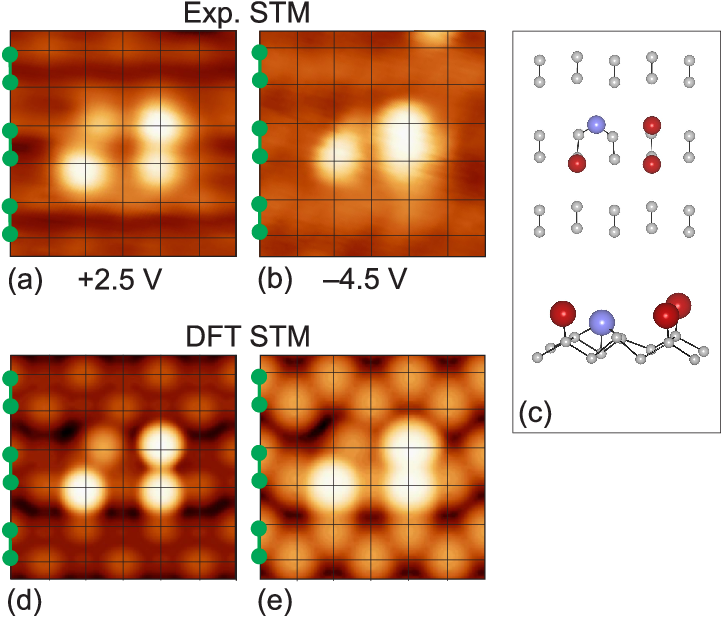}
\caption{\label{ID} Identification of the most frequently observed object on the \ce{PBr3}-dosed Si(100) surface. (a) Empty ($U_s =+2.5$\,V, I$_t$ = 1.0\,nA) and (b) filled ($U_s =-4.5$\,V, I$_t$ = 2.0\,nA) state STM images of the object. Si dimers are marked with green dumbbells. (c) Top and side views of the model of the object (the p-1 structure in SM). Si atoms are shown in grey, Br in red and P in blue. (d) Simulated empty (U$_s =+2.5$\,V) and (e) filled (U$_s = -4.5$\,V) state STM images of the model shown in (c).}
\end{center}
\end{figure}

\begin{figure*}[t]
\includegraphics[width=\linewidth]{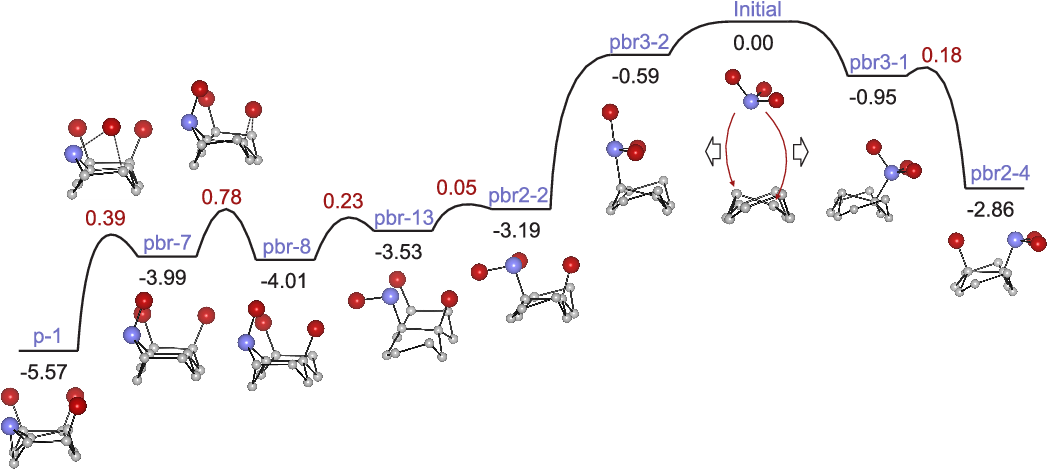}
\caption{\label{react} Energy barrier diagram of \ce{PBr3} dissociative adsorption on the Si(100)-2$\times$1 surface. \ce{PBr3} adsorption on electrophilic Si (`down atom') and nucleophilic Si (`up atom')  are shown to the right and left of the initial state, respectively. Stable structures are shown at the bottom, transition states are shown at the top. The top-view schematics of the stable structures can be found in Fig. S3 in SM. Si atoms are marked in gray, Br in red, and P in blue. All energies are given in electronvolts relative to the initial state energy. Red numbers indicate activation barriers.}
\end{figure*}

To determine the position of protrusions in the most frequently observed object relative to the position of the upper layer of Si(100), we superimposed a grid on the STM images (Fig.~\ref{ID}a, b), where each lattice site corresponds to a Si atom. We found that three protrusions are located above the Si atoms and one protrusion is located in the end-bridge position between the Si atoms of neighboring dimers and above the Si atom of the second layer (Fig.~\ref{ID}a). We associated three protrusions over silicon atoms with bromine atoms. Indeed, Br atoms on the Si(100) surface occupy positions on top of the Si atoms (Fig.~\ref{STM_large}c, d), and it is difficult to observe Br in the bridge position due to fast diffusion \cite{2022PavlovaJChemPhys}. Instead, the end-bridge position is the most preferable for the P atom on the Si(100) surface \cite{2016Warschkow, 1992Brocks, 2006Sen, 2009Bennett}. Therefore, we associated the protrusion between the Si atoms of neighboring dimers with a phosphorus atom. The model  corresponding to the most frequently observed object is shown in Fig.~\ref{ID}c. This structure, p-1, is the most stable among all those calculated (see SM). Notably, the same structure is the most stable in the case of phosphine dissociated on the Si(100) surface \cite{2004Wilson, 2016Warschkow}. The simulated STM images of the p-1 structure (Fig.~\ref{ID}d,e) are in good agreement with the STM images of the most frequently observed object (Fig.~\ref{ID}a,b). Note that only bromine atoms are clearly visible in the filled state STM image (Fig.~\ref{ID}b,e).

In addition to the most frequently observed object, there are other objects on the STM images of the \ce{PBr3}-dosed surface (Fig.~\ref{STM_large}a,b). For the completely dissociated \ce{PBr3} molecule, each protrusion in the STM image can be associated with a P or Br atom based on the successful identification of the most frequently observed object. Indeed, in the STM image at a positive voltage of $+2.5$ V (Fig.~\ref{ID}a), both Br and P atoms are visualized as protrusions, and it is difficult to distinguish them from each other. However, in an STM image of the same object at a negative voltage of $-4.5$ (Fig.~\ref{ID}b), a P atom is almost invisible (or visible as deep depression at $-2.8$ V, see Fig.~S2 in SM), while Br atoms are still visible. Therefore, if  four protrusions are visible in the empty state STM image, and only three protrusions are visible in the filled state STM image, then the fourth protrusion, which is not visible in the filled state STM image, is a P atom. Moreover, atoms can be identified by the adsorption site, because bromine is located above the silicon atom, and phosphorus is in the end-bridge position.

Sometimes brighter objects than bromine atoms were observed on the STM images. In structures with \ce{PBr2} and PBr fragments, the atoms are located higher above the surface than the atoms in a completely dissociated molecule (see Fig. S3 in SM). Preliminarily, we can attribute these bright objects to the incompletely dissociated \ce{PBr3} molecule with \ce{PBr2} and PBr fragments.

\begin{figure}[h]
\begin{center}
\includegraphics[width=\linewidth]{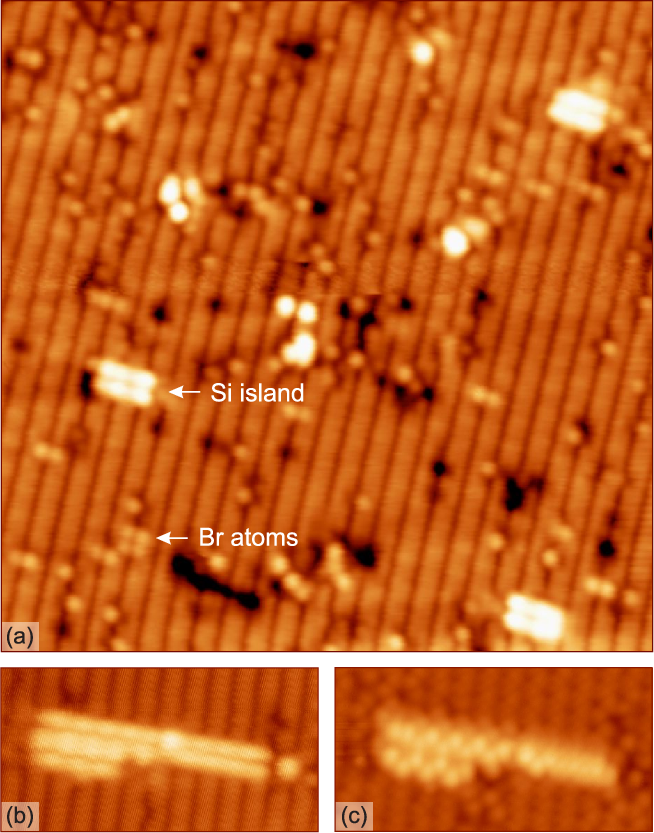}
\caption{\label{STM_Si} The \ce{PBr3}-dosed surface after annealing to 400$^{\circ}$C for 5 min. (a) Empty state STM image (19.3$\times$19.3\,nm$^2$, $U_s =+2.2$\,V, I$_t$ = 2.0\,nA) of the Si(100) surface with silicon islands. (b) Empty ($U_s =+2.5$\,V, I$_t$ = 2.0\,nA) and (c) filled ($U_s =-4.3$\,V, I$_t$ = 2.0\,nA) state STM images of the large silicon island on Si(100). The \ce{PBr3} adsorption was carried out at a partial pressure of $7 \cdot 10^{-10}$\,Torr for ten minutes at room temperature.}
\end{center}
\end{figure}

Figure~\ref{react} shows one of the possible pathways of \ce{PBr3} dissociative adsorption. We calculated activation barriers for elementary stages of \ce{PBr3} dissociation on Si(100)-2$\times$1 with sequential breaking of bonds between P and Br atoms. The detachment of each Br atom from the P atom lowers the total energy of the system by about 1--2\,eV (see SM), resulting in a completely dissociated molecule, which is the most stable structure. The p-1 structure corresponding to the most frequently observed object (Fig.~\ref{ID}) was chosen as the final state of the completely dissociated molecule. In the initial state, \ce{PBr3} was located above the surface with the P atom down, at a P-Si distance of 6\,{\AA}.

We considered two adsorption channels, on electrophilic and nucleophilic Si atoms (Fig.~\ref{react}). Diffusion of the \ce{PBr3} molecule from the nucleophilic atom of the Si dimer (pbr3-2 structure) to the electrophilic atom (pbr3-1 structure) requires an activation energy of 0.39 eV (not shown in Fig.~\ref{react}). Therefore, \ce{PBr3} diffusion inside the Si dimer is unlikely since  \ce{PBr3} can dissociate into  \ce{PBr2} and Br from pbr3-1 (pbr3-2) structures with a lower barrier (without a barrier). Complete dissociation path was calculated for the \ce{PBr3} molecule adsorbed on the nucleophilic Si atom (the left side of the diagram in Fig.~\ref{react}), since in this case \ce{PBr3} spontaneously dissociates into \ce{PBr2} and Br. The highest activation barrier in this dissociation path is 0.78\,eV, which corresponds to Br diffusion between neighboring Si dimers in the same row. At room temperature, the dissociation rate determined by this barrier is 0.1\,s$^{-1}$, i.e. \ce{PBr3} completely dissociates in about 10 seconds. The decomposition of PBr into P and Br occurs significantly faster through the lower reaction barrier of 0.39\,eV ($10^6$ s$^{-1}$ ). Note that we have not considered all possible dissociation pathways, so the existence of another pathway with a lower barrier cannot be ruled out. Therefore, we do not state that the \ce{PBr3} dissociation proceeds along the path indicated in Fig.~\ref{react} through adsorption onto a nucleophilic atom. Moreover, the \ce{PBr3} dissociation occurs along different paths, since different objects were observed in STM images (Fig.~\ref{STM_large}a, b). However, the presence of at least one dissociation path with a barrier that can be crossed with a time scale of about 10 seconds at room temperature confirms that \ce{PBr3} molecules in most cases completely dissociate into P and Br atoms.

The incorporation of a phosphorus atom into silicon requires heating to above room temperature \cite{2003Schofield}. When the sample with \ce{PBr3} adsorbed at room temperature was annealed to 400$^{\circ}$C, islands appeared on the surface (Fig.~\ref{STM_Si}). The islands are formed by Si atoms, since the height of the island is approximately 1.4\, {\AA}, which corresponds to the height of an atomic step on the Si(100) surface. In addition, the filled state STM image of the large island (Fig.~\ref{STM_Si}c) clearly shows buckling, confirming that the islands are formed by Si atoms. The fact that we observed ejection of silicon from the surface indicates the incorporation of phosphorus atoms into silicon \cite{2003Schofield}.

After heating the surface with adsorbed \ce{PBr3}, bromine atoms were observed, which are mainly paired on the Si dimer (Fig.~\ref{STM_Si}a). Bromine monolayer desorbs as \ce{SiBr2} at about 650$^{\circ}$C \cite{1995Flowers}, which is higher than the heating temperature in our case (400$^{\circ}$C), but Br can start to desorb in the form of individual atoms at a lower temperature, 350$^{\circ}$C \cite{2005Trenhaile}. Although the conclusion about Br desorption at 350$^{\circ}$C was made for a saturated bromine coverage \cite{2005Trenhaile}, we cannot exclude the possibility of Br desorption in our case at a low coverage. Nevertheless, some of the bromine atoms definitely remains on the surface, as seen in the STM image in Fig.~\ref{STM_Si}a. To unambiguously identify P atoms, including those in heterodimers, and other defects on the surface, our atomic resolution data are insufficient.

\section{Conclusions}

The \ce{PBr3} adsorption on the Si(100) surface at room temperature is predominantly dissociative, down to individual atoms. The calculation of one of the possible dissociation pathways confirm that the \ce{PBr3} molecule can dissociate on Si(100) at room temperature. In the most frequently observed object on the \ce{PBr3}-dosed surface, bromine are located above the Si atoms, and phosphorus is in the end-bridge position. In most cases, the dissociation of the \ce{PBr3} molecule occurs on three neighboring dimers, as in the case of \ce{PH3} \cite{2004Wilson, 2016Warschkow}. After sample annealing up to 400$^{\circ}$C, the phosphorus incorporates into silicon. Therefore, phosphorus can be incorporated into Si(100) by \ce{PBr3} adsorption, and a window of three dimers in a mask should be sufficient. The obtained results suggest that the \ce{PBr3} molecule can be considered as an alternative to phosphine for near-atomic precision doping of silicon using a halogen monolayer as a mask.

\section{Acknowledgement}

This research was supported in part through computational resources of HPC facilities at HSE University.

\bibliography{PBr3_JPCC_rev_arxiv}

\end{document}